\documentclass[%
 aps,
 physrev,
 groupedaddress,
 amsmath,amssymb,
 floatfix,
]{revtex4-2}

\usepackage{graphicx}
\usepackage{comment}
\usepackage{subcaption}
\usepackage{makecell}
\usepackage{siunitx}
\DeclareSIUnit\angstrom{\text{\AA}}

\usepackage{placeins}
\usepackage{csquotes}
\usepackage{chemformula}
\usepackage{booktabs}

\usepackage{microtype}
\usepackage{orcidlink}

\usepackage{color}
\definecolor{deepblue}{rgb}{0,0,0.5}
\definecolor{deepred}{rgb}{0.6,0,0}
\definecolor{deepgreen}{rgb}{0,0.5,0}
\definecolor{citeGreen}{RGB}{102,204,10}
\definecolor{mypurple}{RGB}{140,54,140}

\usepackage{hyperref}
\hypersetup{
    colorlinks=true,
    linkcolor=blue,
    filecolor=mypurple,
    urlcolor=teal,
    citecolor=citeGreen
}

\newcommand{\eqn}[1]{Eq.~\eqref{#1}}

\begin{document}

\title{Recursive entropy in thermodynamics: the partition function under coarse-graining and the basis of the zentropy approach}

\author{Luke Allen Myers\,\orcidlink{0009-0003-0823-0871}}
 \email{lam7027@psu.edu}
\author{Nigel Lee En Hew\,\orcidlink{0000-0003-1374-4589}}
\author{Shun-Li Shang\,\orcidlink{0000-0002-6524-8897}}
\author{Zi-Kui Liu\,\orcidlink{0000-0003-3346-3696}}
\affiliation{Department of Materials Science and Engineering, The Pennsylvania State University, University Park, Pennsylvania 16802, USA}

\date{\today}

\begin{abstract}
    The recursive property of entropy is well known in information theory;
    however, the concept is underappreciated in thermodynamics,
    the field in which entropy originated.
    The zentropy approach is built on this idea,
    and it has emerged as a useful framework
    for describing thermodynamic systems across multiple scales,
    yet its statistical-physics foundation has not been fully articulated.
    In this work,
    we derive the partition function under coarse-graining
    taking the recursive form of entropy as primary.
    Using probability tree diagrams,
    we first show that the recursive property separates entropy
    into a configurational term and an intra-configurational term,
    and that it reproduces the Gibbs entropy
    whenever each microstate belongs to only one coarse-grained configuration.
    Maximizing entropy in the recursive form yields a partition function
    in which each configuration's Helmholtz energy replaces a microstate energy,
    recovering Rushbrooke's 1940 result for temperature-dependent energy levels
    while maintaining a clear distinction between the physical and statistical aspects of statistical mechanics.
    We argue that the recursive form should be regarded as the primary expression for entropy:
    the Gibbs form is its special case in which every configuration is a microstate,
    and recursivity is itself one of the axioms that uniquely determine the Gibbs entropy.
    Two representative applications, magnetic materials and liquids,
    illustrate how the framework enables physically meaningful coarse-graining
    and captures emergent behavior arising from probability redistribution among configurations.
    Together,
    these results place the zentropy approach on a maximum-entropy foundation
    and provide an intuitive understanding of the coarse-graining of systems.
\end{abstract}

\maketitle

\section{Introduction}
The Gibbs (Shannon) entropy is a function of the probabilities of all possible microstates of a system and quantifies the uncertainty in the state of the system,
\begin{equation} \label{eq:entropy}
    S = -k_B\sum_{x} p_x \ln p_x, 
\end{equation}
where the Boltzmann constant $k_B$ sets the units of entropy.
From this definition,
there is no prescription for the ordering of the microstates.
The only requirement is that the summation be performed over the entire set.
This leaves us with considerable freedom in our \emph{description} of the system.
In fact,
the summation may even be performed hierarchically
with some adjustments,
as is done in the zentropy approach.
Hereafter,
we refer to what is conventionally called a microstate
as a \emph{pure-state configuration},
and to a group of pure-state configurations as a \emph{coarse-grained configuration}.
We use the term \emph{configuration} alone to refer to any description of the system
in which some subset of its degrees of freedom is constrained.
In the present work,
we demonstrate the mathematical foundations and elaborate on the usefulness of the zentropy approach,
where entropy is given by
\begin{equation} \label{eq:zentropy}
    S = - k_B \sum_k p_k \ln p_k + \sum_k p_k S_{k},
\end{equation}
where $p_k$ is the probability of coarse-grained configuration $k$
and $S_k$ is the entropy that remains once the system is known to be in $k$,
which, as an entropy, carries its own factor of $k_B$.
Hereafter we measure entropy in units of $k_B$; that is, we set $k_B = 1$.
At first glance,
there appears to be an extra term $\sum_k p_k S_{k}$; however, \eqn{eq:zentropy} is equivalent to \eqn{eq:entropy} if we group pure-state configurations $x$ into coarse-grained configurations $k$. Furthermore, there is a long history behind \eqn{eq:zentropy}, where it is primarily known as the \textit{recursive property} of entropy \cite{Shannon1948,khinchin1953concept,faddeev1956concept,Aczel1974,Csiszar2008}, which is closely related to the chain rule of entropy \cite{Cover2005}. This property of entropy has been widely used
in the fields of information theory \cite{Cover2005,Bengtsson2017} and quantum mechanics \cite{Wehrl1978,Nielsen2012}.
The zentropy approach is built on the recursive property: it groups pure-state configurations into coarse-grained configurations chosen so that the entropy within each is tractable.

The zentropy approach \cite{Liu2019, Shang2023, Hew2024, Hong2025}
exploits the freedom to choose how pure-state configurations are grouped.
The idea was first applied in the context of density functional theory,
to the $\gamma$--$\alpha$ transition in Ce \cite{Wang2008},
and was later formulated as a method of \enquote{multiscale entropy}
\cite{Liu2019}.
The method subsequently adopted the name \emph{zentropy} \cite{Liu2022},
where \enquote{Z} refers to the partition function $Z$,
itself derived from the German word \emph{Zustandssumme},
meaning \enquote{sum over states}.
The term zentropy emphasizes that this approach sums over states with entropy.

The zentropy approach has been applied in conjunction with spin-polarized density functional theory to incorporate magnetic, electronic, and vibrational degrees of freedom and to predict finite-temperature thermodynamic properties of magnetic materials, with each configuration defined by a distinct atomic spin arrangement. In this manner, zentropy has been used to quantify the degree of disorder \cite{Shang2023},
predict the magnetic phase transitions of bcc Fe \cite{Shang2010a} and fcc Ni \cite{Shang2010}, positive and negative thermal expansion \cite{Liu2022,Liu2024a}, and the ferroelectric phase transition in \ch{PbTiO3} \cite{Hew2024}.
More recently, zentropy has been applied to ab initio molecular dynamics to calculate the entropy of liquids and solids, enabling accurate prediction of melting temperatures \cite{Hong2025,Shang2025}.
Zentropy has also been applied as a framework for machine learning using zentropy-enhanced neural networks (ZENNs) \cite{Wang2025}, and for uncertainty measures for feature selection \cite{Yuan2024a}.

The recursive property and chain rule are fundamental concepts
in the fields of information theory.
They can be found 
at least as far back as 1938 \cite[Chap.~IV, p.~176]{Tolman1938},
and subsequently
immortalized by Shannon in 1948 \cite{Shannon1948}.
For a good basic overview of entropy in information theory, see standard texts \cite{Bengtsson2017,Nielsen2012,Cover2005}.
However,
despite this long history,
the concept is underappreciated in thermodynamics and materials science,
where it can be leveraged as a framework for addressing systems on multiple scales,
so that approximate methods may be applied within configurations where they are accurate.
We show that restating entropy in the form of \eqn{eq:zentropy}
reorganizes the same information into coarse-grained configurations,
each obeying simpler mechanical rules than the system as a whole,
and thereby provides explanatory power for emergent phenomena.

The main contribution of the present work is a derivation
of the partition function under coarse-graining,
obtained by maximizing entropy in its recursive form.
The resulting partition function itself is well established
\cite{Rushbrooke1940,Stillinger1982,Ceder1993,Noid2008,Esposito2012};
what is new is the route.
Because of the recursive property,
a single maximization of entropy applies at every level of the hierarchy,
giving the intra-configurational Helmholtz energy within each coarse-grained configuration
and the partition function across configurations.
To our knowledge,
the partition function has not previously been derived in this way;
Section~\ref{sec:helmholtz} compares this route with earlier derivations,
including a related two-step maximization for a system in contact with a bath \cite{Lee2012}.
The derivation is natural once \eqn{eq:zentropy} is taken as the primary expression for entropy
rather than as a rearrangement of \eqn{eq:entropy},
and we argue that it should be:
\eqn{eq:entropy} is the special case of \eqn{eq:zentropy}
in which every configuration is a pure-state configuration,
and recursivity is one of the axioms from which the Shannon entropy is derived
\cite{khinchin1953concept,faddeev1956concept,Aczel1974}.
Along the way, we provide an intuitive understanding of coarse-graining
aimed at researchers who use, or wish to use, the zentropy approach.

In the remainder of the present work, we will
(i) interpret entropy in terms of constrained and unconstrained degrees of freedom,
which gives the intra-configurational entropy its meaning,
(ii) use probability tree diagrams to show how the recursive property decomposes entropy
and when it reproduces the Gibbs entropy,
(iii) express the Helmholtz energy in terms of coarse-grained configurations
and derive the partition function by maximizing entropy in its recursive form,
and (iv) illustrate, through magnetic materials and liquids,
how coarse-grained configurations are chosen in practice
and how emergent phenomena and the hierarchical application of approximations follow.
We anticipate that the present work
will help researchers understand the zentropy approach,
and will motivate members of the thermodynamics and materials science communities
to make use of the recursive property and zentropy.

\section{The Meaning of Entropy: Constrained and Unconstrained Degrees of Freedom}
For a simple compressible closed system at thermodynamic equilibrium, the fundamental differential relation may be written as
\begin{equation}
    dU = TdS - PdV. 
\end{equation}
In this form, $S$ and $V$ appear symmetrically as conjugate variables to $T$ and $-P$, respectively, and thermodynamics \emph{per se} does not provide insight into the nature of volume and entropy.
Instead, we must rely on mechanics and statistical physics for answers.
While volume is readily understood,
entropy is more mysterious.

To explain entropy, we first note that entropy is an anthropomorphic quantity \cite{Jaynes1965}.
It is a property of the measurement performed on the system.
Entropy does not appear in
Newton's laws of motion, Maxwell's equations, the Schrödinger equation, or Einstein's field equations \cite[Chap.~IV]{Eddington1928}\textsuperscript{,}\cite{Bekenstein1975}.
Instead, it only appears when there are degrees of freedom left unconstrained when performing a measurement.
(While quantum mechanics contains measurement uncertainty due to wavefunction collapse, there is no uncertainty in the state of the wavefunction itself.)
Specifically, entropy quantifies our uncertainty in the state of a system.
It is important to note that
this uncertainty is not due to instrument inaccuracy
but to the unconstrained microscopic degrees of freedom of the system.
When a measurement is performed on a system,
we usually do not fully constrain all its degrees of freedom. 
Instead,
we reduce it to a statistical ensemble of possible configurations
that comport with our measurement.
For example,
the canonical ensemble constrains the number of particles, volume, and temperature,
but does not constrain particle positions.
However,
if we are given some additional constraint,
say the position of a particular particle,
the entropy would be lower.
Hence, the entropy of the system also depends on the constrained degrees of freedom.
Suppose we impose a constraint that specifies a coarse-grained configuration $k$ --- enough
to narrow the ensemble,
but not enough to determine which pure-state configuration the system occupies.
The entropy that remains is then the intra-configurational entropy $S_k$ of \eqn{eq:zentropy},
and Section~\ref{sec:recursive} shows how the $S_k$ combine
to give the entropy of the unconstrained system.

\section{The Recursive Property of Entropy}\label{sec:recursive}
\begin{figure}[h]
    \centering
    \includegraphics[width=1\linewidth]{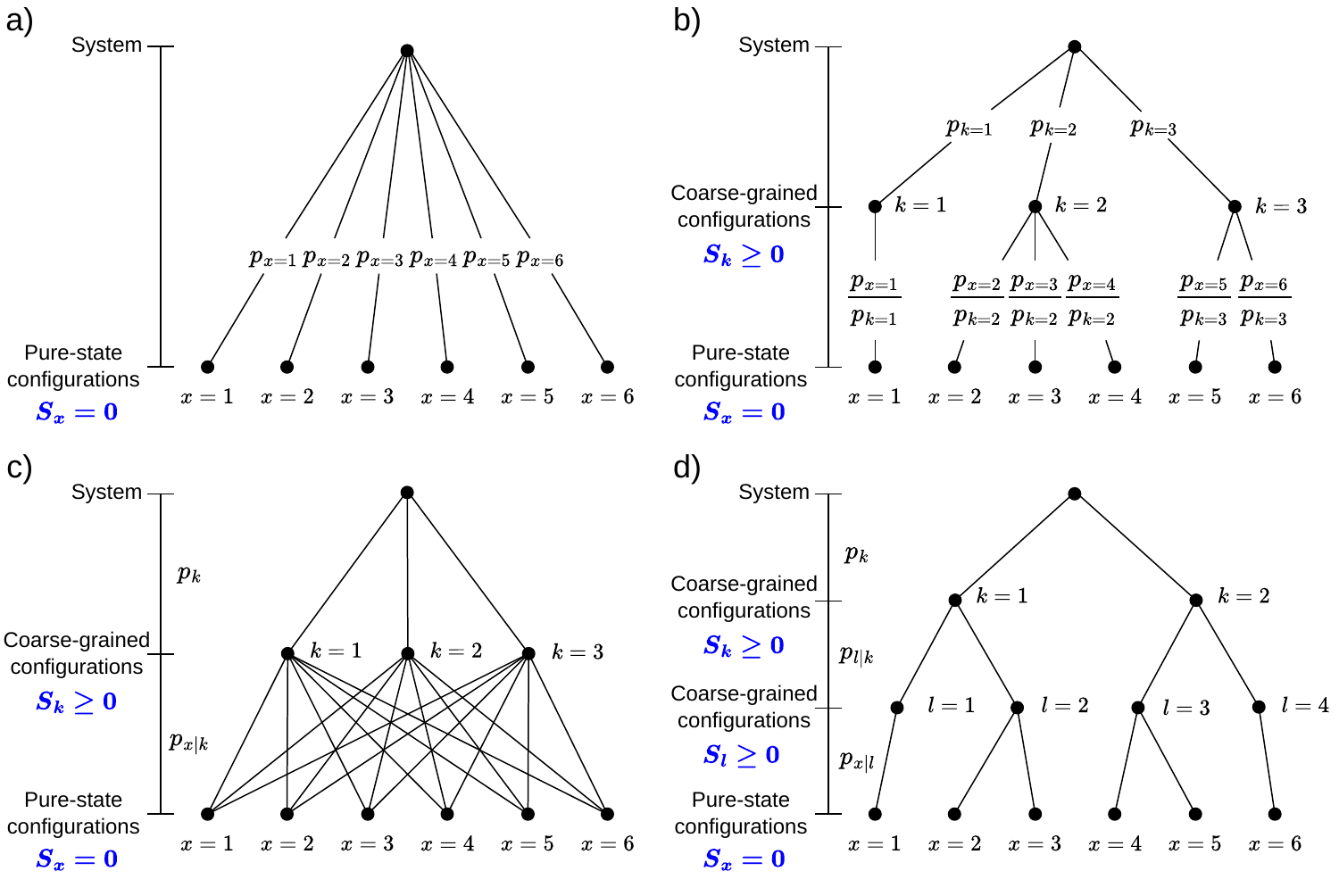}
    \caption{
        Probability tree diagrams for four descriptions of the same system.
        A pure-state configuration $x$ specifies all degrees of freedom ($S_x = 0$),
        whereas a coarse-grained configuration $k$ specifies only some,
        leaving an intra-configurational entropy $S_k \geq 0$.
        (a) No coarse-graining.
        (b) Partitioned coarse-graining: each $x$ belongs to exactly one $k$.
        (c) General coarse-graining: an $x$ may be reached through several $k$,
        which adds a labeling entropy $S_{\text{label}}$ (\eqn{eq:S_label}) that vanishes in (b).
        (d) Recursive coarse-graining: each $k$ is itself partitioned into finer coarse-grained configurations $l$.
        In (c) and (d), the probability carried by the branches between two levels
        is labeled once beside the vertical axis rather than on each branch.
    }
    \label{fig:recursion_tree}
\end{figure}

The recursive property, \eqn{eq:zentropy}, is most easily seen
using probability tree diagrams \cite[Chap.~2]{Bengtsson2017},
as shown in Figure~\ref{fig:recursion_tree}.
Without coarse-graining (Figure~\ref{fig:recursion_tree}a),
each branch leads from the system directly to a pure-state configuration $x$
with probability $p_x$,
and the entropy is given by \eqn{eq:entropy}.
Coarse-graining inserts an intermediate level of coarse-grained configurations $k$
(Figure~\ref{fig:recursion_tree}b,c).
Each branch from the system to a coarse-grained configuration $k$ carries the probability $p_k$.
Each branch from $k$ to a pure-state configuration $x$ carries the conditional probability $p_{x \mid k}$,
and the path through both carries the product
\begin{equation}\label{eq:joint}
    p_{x,k} = p_k\, p_{x \mid k}.
\end{equation}
Summing over paths recovers the probabilities at each level,
$p_k = \sum_x p_{x,k}$ and $p_x = \sum_k p_{x,k}$.
The entropy that remains once the system is known to be in $k$
is the \emph{intra-configurational} entropy
\begin{equation}\label{eq:Sk}
    S_k = -\sum_x p_{x \mid k} \ln p_{x \mid k},
\end{equation}
written here directly over pure-state configurations;
its general form is given in \eqn{eq:Sk_recursive}.

Consider the entropy $S$ of the distribution over paths.
Factoring each path probability using \eqn{eq:joint}
and applying $\ln ab = \ln a + \ln b$,
\begin{align*}
    S &= -\sum_k \sum_x p_{x,k} \ln p_{x,k} \\
      &= -\sum_k \sum_x p_k\, p_{x \mid k} \left( \ln p_k + \ln p_{x \mid k} \right) \\
      &= -\sum_k p_k \ln p_k \sum_x p_{x \mid k}
         - \sum_k p_k \sum_x p_{x \mid k} \ln p_{x \mid k} \\
      &= \underbrace{-\sum_k p_k \ln p_k}_{S_{\text{conf}}}
         + \underbrace{\sum_k p_k S_k}_{\langle S_k \rangle},
\end{align*}
where the last step uses $\sum_x p_{x \mid k} = 1$ and \eqn{eq:Sk}.
This is \eqn{eq:zentropy}.
The first term is the \emph{configurational} entropy
\cite{AdamGibbs1965,MiracleSenkov2017,Hong2025,Shang2023},
the entropy of the distribution over coarse-grained configurations,
and the second is the expectation value of the intra-configurational entropy.

In the partitioned case (Figure~\ref{fig:recursion_tree}b),
each pure-state configuration belongs to exactly one coarse-grained configuration.
Writing $G_k$ for the set of pure-state configurations in $k$,
we have $p_k = \sum_{x \in G_k} p_x$
and $p_{x \mid k} = p_x / p_k$ for $x \in G_k$ and zero otherwise.
Each pure-state configuration then lies on a single path with $p_{x,k} = p_x$,
so $S$ is exactly \eqn{eq:entropy},
and \eqn{eq:zentropy} simply sums the same terms group by group.
For this reason,
the recursive property is also called the \emph{grouping} property \cite{Cover2005}.

In general (Figure~\ref{fig:recursion_tree}c),
a pure-state configuration may be reached through more than one coarse-grained configuration.
The derivation above still holds,
since it relies only on \eqn{eq:joint},
but $S$ is now the entropy of the joint distribution $p_{x,k}$
rather than of the pure-state configurations alone.
Factoring each path probability instead as $p_{x,k} = p_x \left( p_{x,k}/p_x \right)$
separates the two:
\begin{equation}\label{eq:S_label}
    S = -\sum_x p_x \ln p_x
        \underbrace{- \sum_x \sum_k p_{x,k} \ln \frac{p_{x,k}}{p_x}}_{S_{\text{label}}}.
\end{equation}
The first term is \eqn{eq:entropy}.
The second, $S_{\text{label}}$, is the uncertainty in which coarse-grained configuration
is assigned to a given pure-state configuration.
It is non-negative, since $p_{x,k} \leq p_x$,
and vanishes in the partitioned case,
where $p_{x,k} = p_x$ on the single path to each pure-state configuration.
However,
$S_{\text{label}}$ is not an entropy of the system.
Because a pure-state configuration specifies every degree of freedom,
if all coarse-grained configurations are defined by a property of the system,
such as a spin arrangement, a coordination environment, or a basin of the energy landscape,
then each coarse-grained configuration $k$ is fixed by $x$,
and $S_{\text{label}} = 0$.
A nonzero $S_{\text{label}}$ arises only when $k$ is a label assigned to the system
rather than one of its degrees of freedom,
so the partitioned case is the one of physical interest.
Ceder identified the same requirement for the Ising description of alloys \cite{Ceder1993}.

The property is recursive because each $S_k$ is itself an entropy
and may be decomposed in the same way (Figure~\ref{fig:recursion_tree}d).
Introducing an intermediate level of coarse-grained configurations $l$,
each belonging to exactly one $k$,
\begin{equation}\label{eq:Sk_recursive}
    S_k = -\sum_l p_{l \mid k} \ln p_{l \mid k} + \sum_l p_{l \mid k} S_l,
\end{equation}
where $S_l$ is the entropy that remains once $l$ is specified;
because each $l$ lies within a single $k$, specifying $l$ also specifies $k$.
\eqn{eq:Sk_recursive} has the same form as \eqn{eq:zentropy},
with the system replaced by $k$ and $k$ replaced by $l$,
and it is the general form of the intra-configurational entropy:
when each $l$ is a pure-state configuration, $S_l = S_x = 0$,
and \eqn{eq:Sk_recursive} reduces to \eqn{eq:Sk},
just as \eqn{eq:zentropy} reduces to \eqn{eq:entropy}
when every coarse-grained configuration is a pure-state configuration.
The hierarchy may therefore be extended to any depth,
with the entropy at every level taking the form of \eqn{eq:zentropy},
and terminates at the pure-state configurations.

In the random-variable formulation, as commonly employed in information theory,
\eqn{eq:zentropy} is closely related to the chain rule $S(K,X) = S(K) + S(X \mid K)$
\cite{Cover2005,Bengtsson2017,Nielsen2012},
with $S_{\text{conf}}$ corresponding to the entropy $S(K)$ of the coarse-grained description
and $\langle S_k \rangle$ to the conditional entropy $S(X \mid K)$.
Expanding the same joint entropy in the other order,
$S(K,X) = S(X) + S(K \mid X)$,
identifies $S_{\text{label}}$ in \eqn{eq:S_label} as the conditional entropy $S(K \mid X)$.
Comparing the two expansions gives the mutual information
$I(X;K) = S(X) - S(X \mid K) = S_{\text{conf}} - S_{\text{label}}$,
the information that specifying the coarse-grained configuration
provides about the pure-state configuration.
In the partitioned case, $S_{\text{label}} = 0$ and $S_{\text{conf}} = I(X;K)$:
the configurational entropy is exactly the information gained by learning $k$,
and $\langle S_k \rangle$ is the uncertainty that remains.
The chain rule, however, treats the conditional entropy as a single quantity,
whereas \eqn{eq:zentropy} writes it explicitly
as an average of the entropies $S_k$ of the individual coarse-grained configurations,
each of which is an entropy in its own right and may be decomposed again.
It is this form that makes the recursivity of entropy explicit.
The two also differ in logical status:
the chain rule is derived from the Shannon entropy,
whereas the recursive property, together with a few other natural requirements,
may be taken as an axiom from which the Shannon entropy is uniquely determined
\cite{khinchin1953concept,faddeev1956concept,Aczel1974}.

Because \eqn{eq:zentropy} is exact
for any coarse-graining in which each pure-state configuration
belongs to exactly one coarse-grained configuration,
the decomposition itself introduces no approximation.
Any approximation enters only later,
when the intra-configurational entropies $S_k$ are evaluated.
The coarse-grained configurations should therefore be chosen
so that each is simple enough for standard methods
to evaluate its $S_k$ accurately,
even where those methods would fail for the system as a whole.
This freedom to choose the scale of description
is what makes the recursive property useful for systems on multiple scales.

\section{Helmholtz Energy and the Partition Function}\label{sec:helmholtz}

The Helmholtz energy and the partition function are defined downstream of entropy,
so evaluating them for coarse-grained configurations requires accounting for
intra-configurational entropy.
Because a coarse-grained configuration carries both energy and entropy,
it may be regarded as a sub-ensemble.
In 1940, Rushbrooke described such systems as the
\enquote{statistical mechanics of assemblies whose energy-levels depend on the temperature}~\cite{Rushbrooke1940},
and showed that the consequence is simply a substitution:
an intra-configurational Helmholtz energy $F_k$ takes the place of the energy $E_x$
in both the Helmholtz energy and the partition function.
Noting that the result appears in earlier treatments by Fowler and Guggenheim,
he derived it explicitly for a simple class of assemblies
by summing the partition function over the degrees of freedom within each energy level,
and argued for its general validity from consistency with thermodynamics.
The same substitution underlies inherent-structure descriptions of liquids \cite{Stillinger1982},
the derivation of the Ising model for alloy phase diagrams \cite{Ceder1993},
and potentials of mean force in molecular coarse-graining \cite{Noid2008};
in each case it is obtained by splitting the sum over microstates
into restricted partition functions, one per coarse-grained configuration.
Esposito reached the same result from stochastic dynamics \cite{Esposito2012}:
when transitions within each coarse-grained configuration are much faster than those between them,
the master equation closes at the coarse-grained level,
and its stationary distribution weights each configuration
by the exponential of its free energy.
Each of these treatments assumes that every microstate belongs to exactly one coarse-grained configuration,
consistent with Section~\ref{sec:recursive}.

None of these derivations maximizes entropy written in recursive form.
The partition-function route starts from the ordinary partition function,
itself obtainable by maximizing entropy over pure-state configurations \cite{Jaynes1957,Jaynes1957a},
and the dynamical route builds the Boltzmann weights into the transition rates
through local detailed balance.
Lee and Press\'e did maximize an entropy in recursive form, in two steps,
but for a different purpose \cite{Lee2012}.
They considered an open system and its bath at fixed total energy,
so that the inner maximization, over bath states for each state of the system, is microcanonical,
and the outer one takes the canonical form only once the bath is assumed large.
Their result is the ordinary Boltzmann distribution over the states of the system,
and it shows why contact with a bath at fixed temperature
is equivalent to constraining the mean energy of the system.
Here, we take that mean-energy constraint as given,
and the coarse-grained configurations are descriptions of the system itself,
each with its own intra-configurational entropy.
Because of the recursive property,
the inner and outer maximizations are then the same problem:
within each coarse-grained configuration it yields the intra-configurational Helmholtz energy,
and across configurations it yields the partition function.
To the best of our knowledge,
the partition function has not previously been derived in this way.
We present that derivation here,
taking the recursive property as the primary equation for entropy.
In this form,
each coarse-grained configuration enters the partition function through its Helmholtz energy $F_k$,
just as a pure-state configuration enters through its energy $E_x$,
and Jaynes' sharp distinction between the physical and statistical aspects
of statistical mechanics is maintained.

From thermodynamics, the \emph{Helmholtz energy} is defined as
\begin{equation} \label{eq:helmholtz}
    F = U - TS.
\end{equation}
Since $x$ specifies all degrees of freedom,
each pure-state configuration has a definite energy $E_x$,
independent of the coarse-graining $k$.
The \textit{internal energy} $U$ of the thermodynamic system is then the expectation value
of the energy over the pure-state configurations $x$:
\begin{equation}
    U = \langle E \rangle = \sum_{x} p_x E_{x} = \sum_{k}\sum_{x} p_{x,k} E_{x},
\end{equation}
where the second equality follows from $p_x = \sum_k p_{x,k}$.
Just as $S_k$ is the entropy of the conditional distribution $p_{x \mid k}$ (\eqn{eq:Sk}),
the internal energy of a coarse-grained configuration $k$
is the expectation of the energy over that distribution:
\begin{equation}
    U_k = \langle E \rangle_k = \sum_{x} p_{x \mid k} E_{x}.
\end{equation}
By the law of total expectation---the expectation-value counterpart of the law of total
probability, valid for any joint distribution---the system energy is recovered as the
expectation of $U_k$ over the coarse-grained configurations:
\begin{equation}\label{eq:total_energy}
    U = \langle U_k \rangle = \sum_k p_k U_k.
\end{equation}
Substituting \eqn{eq:total_energy} and the recursive property \eqn{eq:zentropy}
into the definition of the Helmholtz energy (\eqn{eq:helmholtz}) gives
\begin{equation}\label{eq:f_of_ek_sk}
    F = \sum_k p_k U_k + T \left( \sum_k p_k \ln p_k - \sum_k p_k S_k \right).
\end{equation}
Defining the intra-configurational Helmholtz energy $F_k = U_k - T S_k$, \eqn{eq:f_of_ek_sk} simplifies to
\begin{equation}\label{eq:helmholtz_par}
    \boxed{F = \sum_k p_k F_k + T \sum_k p_k \ln p_k.}
\end{equation}
The intra-configurational Helmholtz energy $F_k$ thus replaces the energy $E_{x}$ of the
pure-state configurations.
Because both the recursive property and the law of total expectation are exact,
this decomposition holds for any coarse-graining
in which each pure-state configuration belongs to exactly one coarse-grained configuration
(Figure~\ref{fig:recursion_tree}b).

Next,
we derive the partition function for the canonical ensemble
by maximizing entropy in its recursive form (\eqn{eq:zentropy})
by optimizing the probabilities $p_k$ of the coarse-grained configurations $k$
under the constraints that the sum of the probabilities is equal to one,
\begin{equation} \label{eq:probability_constraint}
    \sum_k p_k = 1,
\end{equation}
and that the expectation value of the energy is a constant ($U=\mathrm{const.}$).
Throughout the variation, $U_k$ and $S_k$ are treated as fixed for each configuration $k$.
This is not a further constraint on the distribution but a statement about the order in which the
optimization is carried out: the distributions internal to each configuration are optimized first,
by the same maximization applied one level down, as shown below.
Following the method of Lagrange multipliers in the calculus of variations, the Lagrange function is
\begin{equation}
    \mathcal{L} = \left(-\sum_k p_k \ln p_k + \sum_k p_k S_k \right)
    + \alpha \left( 1 - \sum_k p_k \right)
    + \beta \left( U - \sum_k p_k U_k \right).
\end{equation}
To extremize entropy, we set the variation of the Lagrange function to zero and simplify:
\begin{align}
    0 &= \delta\mathcal{L} \\
      &= \delta \left(-\sum_k p_k \ln p_k + \sum_k p_k S_k \right)
      + \delta \left( \alpha - \sum_k \alpha p_k \right)
      + \delta \left( \beta U - \sum_k \beta p_k U_k \right) \\
      &= \sum_k \left[
            \delta \left(-p_k \ln p_k + p_k S_k \right)
            - \delta \left( \alpha p_k \right)
            - \delta \left( \beta U_k p_k \right)
        \right] \\
      &= \sum_k \left[
            \frac{\partial}{\partial p_k} \left(-p_k \ln p_k \right)  \delta p_k
            + \frac{\partial}{\partial p_k} \left(p_k S_k \right)\delta p_k
            - \frac{\partial}{\partial p_k} \left( \alpha p_k \right) \delta p_k
            - \frac{\partial}{\partial p_k} \left( \beta U_k p_k \right) \delta p_k
        \right] \\
      &= \sum_k \left[ -\ln p_k -1 + S_k - \alpha - \beta U_k\right] \delta p_k.
\end{align}
Since the variations are independent, the coefficient of each $\delta p_k$ must vanish. Thus, for every $k$,
\begin{equation}
0 = -\ln p_k - 1 + S_k - \alpha - \beta U_k.
\end{equation}
Isolating $p_k$ yields
\begin{equation} \label{eq:pk_of_alpha_beta}
    p_k = e^{\left( -1  -\alpha -\beta U_k + S_k \right)} = e^{-(1 + \alpha)}\, e^{\left(-\beta U_k + S_k \right)}.
\end{equation}
Applying the probability constraint (\eqn{eq:probability_constraint}),
\begin{equation}\label{eq:alpha_to_beta}
    1 = \sum_k p_k = e^{-(1 + \alpha)} \sum_k e^{\left(-\beta U_k + S_k \right)}.
\end{equation}
The partition function is then naturally defined as
\begin{equation}\label{eq:def_z}
    Z = e^{\left( 1 + \alpha \right)},
\end{equation}
so that \eqn{eq:alpha_to_beta} gives the relation between the partition function and $\beta$,
\begin{equation} \label{eq:z_of_beta}
    Z = \sum_k e^{\left(-\beta U_k + S_k \right)},
\end{equation}
and \eqn{eq:pk_of_alpha_beta} gives the configuration probabilities
\begin{equation}\label{eq:p_k}
    p_k = \frac{1}{Z} e^{\left(-\beta U_k + S_k \right)}.
\end{equation}

Before confirming $\beta$ as the inverse temperature, we justify treating $U_k$ and $S_k$ as fixed during the variation over $p_k$.
At fixed $\beta$, maximizing the entropy subject to the energy constraint is equivalent to
maximizing
\begin{equation}\label{eq:inner_outer}
    S - \beta U = -\sum_k p_k \ln p_k + \sum_k p_k \left( S_k - \beta U_k \right).
\end{equation}
The quantity $S_k - \beta U_k$ is a functional of the conditional distribution $p_{x \mid k}$ alone,
and it enters only through the non-negative weight $p_k$.
Maximizing each $S_k - \beta U_k$ therefore maximizes \eqn{eq:inner_outer} whatever the $p_k$ may be,
so the resulting $U_k$ and $S_k$ depend on $\beta$ but not on $p_k$ --- which is what the variation
above assumes.
This is the subset-independence property that Shore and Johnson showed
any consistent maximum-entropy inference must satisfy \cite{Shore1980}:
conditional distributions within disjoint subsets of states
may be inferred independently of the probabilities of the subsets.

Because of the recursive property,
this maximization takes the same form at every level of the hierarchy.
Within a coarse-grained configuration $k$,
$S_k$ is itself an entropy of the form of \eqn{eq:zentropy} over finer configurations (\eqn{eq:Sk_recursive}),
so maximizing $S_k - \beta U_k$ is the same problem as maximizing \eqn{eq:inner_outer},
one level down,
with the same $\beta$, since there is only one constraint, on the total mean energy.
The hierarchy terminates at the pure-state configurations,
which have definite energies and no intra-configurational entropy ($U_x = E_x$, $S_x = 0$).
Applying \eqn{eq:p_k} at this level gives
\begin{equation}\label{eq:inner_step}
    p_{x \mid k} = \frac{e^{-\beta E_x}}{Z_k},
    \qquad
    Z_k = \sum_{x \in G_k} e^{-\beta E_x},
\end{equation}
so that
\begin{equation}\label{eq:Fk_lnZk}
    F_k = U_k - T S_k = -T \ln Z_k .
\end{equation}
The intra-configurational Helmholtz energy is thus itself the logarithm of a partition function,
restricted to configuration $k$,
as in earlier work \cite{Ceder1993,Noid2008},
and the partition function has the same form at every level,
with $F_x = E_x$ for pure-state configurations.
In the partitioned case,
$Z = \sum_k e^{-\beta F_k} = \sum_k Z_k = \sum_x e^{-\beta E_x}$,
the ordinary partition function.

We can now confirm that the identity of $\beta$ is indeed inverse temperature
and is unaffected by the inclusion of intra-configurational entropy.
Rewriting the recursive property (\eqn{eq:zentropy}) using \eqn{eq:p_k} to evaluate $\ln p_k$ (leaving the prefactor and other instances of $p_k$ unchanged):
\begin{align}
    S &= -\sum_k p_k \ln \left[\frac{1}{Z} e^{\left(-\beta U_k + S_k \right)}\right] + \sum_k p_k S_k \\
      &= - \sum_k p_k \left( -\beta U_k + S_k - \ln{Z} \right)+ \sum_k p_k S_k  \\
      &= - \sum_k p_k \left( -\beta U_k + S_k - \ln{Z} - S_k  \right) \label{eq:before_cancel} \\
      &= - \sum_k p_k \left( -\beta U_k - \ln{Z}  \right) \\
      &= - \sum_k \left( -p_k \beta U_k - p_k \ln{Z} \right) \\
      &= \beta \sum_k p_k  U_k + \ln{Z} \sum_k p_k \\
      &= \beta U + \ln{Z}. \label{eq:s_of_beta_z}
\end{align}
Notice that the $S_k$ terms cancel out in \eqn{eq:before_cancel}.
Differentiating \eqn{eq:s_of_beta_z} with respect to the internal energy, and treating $\beta$ as a
function of $U$, gives
\begin{equation}\label{eq:dS_dU}
    \left(\frac{\partial S}{\partial U}\right)_{V,N}
    = \beta + \left( U + \frac{\partial \ln Z}{\partial \beta} \right)
      \left(\frac{\partial \beta}{\partial U}\right)_{V,N}.
\end{equation}
The bracketed term vanishes, as we now show.
Although $U_k$ and $S_k$ are independent of $p_k$, they do depend on $\beta$,
and that dependence is constrained:
because each $S_k - \beta U_k$ is stationary with respect to its own internal distribution,
the envelope theorem gives
\begin{equation}\label{eq:envelope}
    \frac{\partial S_k}{\partial \beta} = \beta \frac{\partial U_k}{\partial \beta}.
\end{equation}
Differentiating \eqn{eq:z_of_beta} then gives
\begin{equation}
    \frac{\partial \ln Z}{\partial \beta}
    = \sum_k p_k \left( -U_k - \beta \frac{\partial U_k}{\partial \beta}
      + \frac{\partial S_k}{\partial \beta} \right)
    = -\sum_k p_k U_k = -U,
\end{equation}
where the derivatives of $U_k$ and $S_k$ cancel by the envelope relation \eqn{eq:envelope}.
\eqn{eq:dS_dU} therefore reduces to
\begin{equation}
    \left(\frac{\partial S}{\partial U}\right)_{V,N} = \beta,
\end{equation}
and by the definition of temperature $T = \left(\frac{\partial U}{\partial S}\right)_{V,N}$,
\begin{equation}\label{eq:beta}
    \beta = \frac{1}{T},
\end{equation}
as expected.

With $\beta = 1/T$ established, the partition function (\eqn{eq:z_of_beta}) may be written in terms
of the intra-configurational Helmholtz energy $F_k = U_k - T S_k$,
\begin{equation}\label{eq:partition_function}
    \boxed{Z = \sum_k e^{\left(-\frac{F_k}{T} \right)} = \sum_k e^{\left(-\beta F_k \right)}.}
\end{equation}
Note that while the definition of the partition function (\eqn{eq:def_z}) and the identity of $\beta$ (\eqn{eq:beta}) are the same as in the standard thermodynamic approach,
the derived relation between the partition function and $\beta$ (\eqn{eq:partition_function}) is not.
The difference is that the energy $E_x$ of each pure-state configuration
is replaced by the intra-configurational Helmholtz energy $F_k$ of each coarse-grained configuration,
as expected.
Finally,
since $S = \beta U + \ln Z$ (\eqn{eq:s_of_beta_z}) with $\beta = 1/T$,
the equilibrium Helmholtz energy takes the familiar form
\begin{equation}\label{eq:F_lnZ}
    F = U - TS = -T \ln Z,
\end{equation}
now with $Z$ given by \eqn{eq:partition_function}.

\section{Usefulness of the Zentropy Approach}

Why do we need recursive entropy (\eqn{eq:zentropy}) when we have the standard definition (\eqn{eq:entropy})? 
Why group pure-state configurations at all?
The answer is that
recursive entropy is a generalization of the form of the standard definition
that lends itself to accurate application of approximate methods,
by confining each approximation to a configuration in which it holds
even where it would fail for the system as a whole.
This applies to any system
whose pure-state configurations
may be partitioned into basins.
This partitioning into basins can be used to explain emergent phenomena
in which the system exhibits behaviors absent in the individual configurations.
Thus, a framework
such as zentropy,
which incorporates a hierarchical partitioning of pure-state configurations,
offers a natural explanation for emergent phenomena.
For example,
the zentropy approach has been used to predict negative thermal expansion \cite{Liu2022,Liu2024a},
magnetic phase transitions \cite{Shang2010,Shang2010a,Shang2023},
and ferroelectric phase transitions \cite{Hew2024}.

For any partition of the pure-state configurations,
\eqn{eq:zentropy} gives the same entropy as \eqn{eq:entropy}
(Section~\ref{sec:recursive});
what the choice of grouping changes is how that entropy is divided between the two terms.
At one extreme, the grouping is as fine as possible:
each coarse-grained configuration $k$ contains exactly one pure-state configuration
(the number of $k$ is maximal),
so $S_k = 0$ for all $k$ and only the first term $-\sum_k p_k\ln p_k$ remains.
Identifying $k = x$,
\eqn{eq:zentropy} then takes the form of \eqn{eq:entropy} term by term.

Conversely, at the opposite extreme, the grouping may be made as coarse as possible,
so that a single coarse-grained configuration contains every pure-state configuration
(the number of $k$ is minimal), in which case the configurational entropy term vanishes
($-\sum_k p_k \ln p_k = 0$)
and \eqn{eq:zentropy} becomes trivial
$S = \sum_k p_k S_k = S_k$.
In practice, however, the most useful description typically lies between these two extremes.
All this is to say that including the intra-configurational entropy term $\sum_k p_k S_k$ allows us to choose a definition of configuration that is best suited to the problem at hand.
Or, in other words,
to define the configurations of the system
at the most useful scale. We emphasize that the recursive property is not a statement about the physical nature of a system; rather, its usefulness lies in the flexibility it gives us in describing the system in terms of coarse-grained configurations.

A useful way to group pure-state configurations
is the way nature already partitions them:
each basin of the energy landscape becomes one coarse-grained configuration.
It is then natural to consider the system's behavior within a given basin.
This can be used to simplify calculations
by applying approximations hierarchically,
as the pure-state configurations in each basin
often have shared properties, such as harmonic behavior,
that may be absent from the system as a whole.

To concretely show the usefulness of the zentropy approach,
we consider two applications:
magnetic materials and liquids.
These examples are illustrative rather than exhaustive.
Because the recursive property is a general feature of entropy,
the zentropy framework applies wherever a statistical ensemble admits a natural hierarchical grouping,
whether by energy basin, by coordination environment, or by some other choice suited to the problem.
Recent work has extended the approach to machine learning \cite{Wang2025,Yuan2024a},
and the same framework can in principle be applied to any problem in statistical mechanics.

\subsection{Magnetic Materials}

The zentropy approach has been applied to magnetic materials \cite{Shang2023,Liu2022,Shang2010,Shang2010a,Liu2024},
with the assumption that the magnetic atoms 
can be either spin-up or spin-down.
Here, the natural grouping is by magnetic configuration:
each coarse-grained configuration $k$ is one arrangement
of the magnetic moments over the atoms,
and the pure-state configurations grouped into $k$
are those sharing that arrangement.
For a given magnetic configuration $k$,
the remaining unconstrained degrees of freedom
are the electronic and vibrational degrees of freedom.
If we assume that these degrees of freedom are independent \cite{Wang2004},
the entropies are additive and the intra-configurational entropy of $k$ is
\begin{equation} \label{eq:el_plus_vib}
    S_k =  S_{k,\text{el}} + S_{k,\text{vib}}.
\end{equation}
Practically,
the electronic and vibrational contributions can be calculated
using density functional theory.
The electronic contribution is obtained by applying Fermi–Dirac statistics
to the electronic density of states,
while the vibrational contribution is obtained
by applying Bose–Einstein statistics to the phonon density of states
or, alternatively,
by using the Debye–Grüneisen model \cite{Liu2015}.

This grouping works because each magnetic configuration labels a distinct basin
of the energy landscape.
The coordinates of that landscape are the atomic positions and the volume,
with the arrangement of magnetic moments entering as a discrete label:
spin-polarized density functional theory gives, for each $k$,
a potential-energy surface over those coordinates with its own minimum,
and hence its own equilibrium volume, phonon spectrum,
and electronic density of states.
The basin of $k$ is the neighborhood of that minimum.
Within it,
atomic displacements are small and approximately harmonic,
and the electronic structure is fixed by the spin arrangement,
so the electronic and vibrational contributions
can be treated nearly independently.
Across basins, in contrast,
these same properties can change appreciably with the magnetic arrangement.
Magnetic configurations therefore form a strongly lumpable partition \cite{Rosas2024}:
each is identified by a set of physical properties
clearly distinguishable from those of its neighbors,
which is the regime in which the recursive property is most informative.

Within this framework,
thermodynamic anomalies emerge directly from the temperature dependence
of the probabilities $p_k = e^{-\beta F_k}/Z$ (Eqs.~\eqref{eq:p_k} and~\eqref{eq:partition_function}).
Consider the Invar alloy \ch{Fe3Pt} \cite{Shang2023,Liu2022,Hew2026}.
Its ferromagnetic ground state
happens to occupy the largest equilibrium volume of any magnetic arrangement.
At low temperatures,
the $-TS_k$ term contributes little to each $F_k$,
so the ferromagnetic configuration, having the lowest energy, also has the lowest $F_k$;
the ensemble sits almost entirely in the ferromagnetic basin,
and the alloy takes on that larger volume.
As temperature increases,
the energy differences between configurations become small compared with $T$,
and the $-TS_k$ term in each $F_k$ grows in importance,
so the Boltzmann factors $e^{-\beta F_k}$ of the many ferrimagnetic configurations,
which have higher energies but smaller equilibrium volumes,
grow relative to that of the ferromagnetic configuration.
Probability flows from the ferromagnetic configuration into the ferrimagnetic ones,
and the ensemble-averaged volume $V = \sum_k p_k V_k$ contracts,
where $V_k$ is the equilibrium volume of configuration $k$.
Negative thermal expansion is recovered as a direct consequence,
with no fitted parameters. Analogous treatments have been applied to
body-centered cubic Fe \cite{Shang2010a} and face-centered cubic Ni \cite{Shang2010}.

In these examples, we see that,
within any single magnetic configuration,
atomic displacements are approximately harmonic
and physical properties vary only smoothly with temperature.
The negative thermal expansion and other anomalies are properties of the system but not any individual $k$;
each emerges only from the temperature-driven redistribution of probability
across configurations.
These properties are emergent:
the rules governing a single basin need not govern the ensemble,
and harmonic behavior within configurations
can give rise to qualitatively new behavior at the macroscopic scale.

\subsection{Liquids}
More recently,
the zentropy approach was applied
to the calculation of entropy in liquids using molecular dynamics \cite{Hong2025},
where it offers an alternative method for calculating the configurational entropy.
Here, the natural grouping is by local coordination environment:
each coarse-grained configuration $k$ is one such environment defined by
the type and number of nearest neighbors of an atom,
and the pure-state configurations grouped into $k$
are those sharing that environment.
This grouping is justified by a separation of timescales.
Within a fixed coordination environment,
atoms oscillate rapidly about their instantaneous equilibrium positions,
exploring vibrational degrees of freedom on timescales much shorter
than those on which the coordination itself changes.
The coordination environment is therefore effectively a slow variable,
while the vibrational and electronic degrees of freedom are fast.
While the bulk of the configurational entropy is accounted for by first-nearest-neighbors,
including the $n$th-nearest-neighbors captures progressively more of it
at the cost of a rapidly growing number of distinct configurations.
For a given coordination environment $k$,
the remaining unconstrained degrees of freedom
are the vibrational and electronic degrees of freedom.
Assuming these degrees of freedom are independent,
the intra-configurational entropy takes the same additive form as in the magnetic case, \eqn{eq:el_plus_vib}.

However,
the practice differs:
instead of calculating the entropy contribution of each configuration individually,
the configurational, vibrational, and electronic entropies are calculated from a single trajectory and then summed:
\begin{equation}
    S = S_\text{conf} + S_{\text{el}} + S_{\text{vib}}.
\end{equation}
Here $S_\text{conf}$ is the configurational term of \eqn{eq:zentropy},
and $S_{\text{el}} + S_{\text{vib}}$ estimates the intra-configurational term $\sum_k p_k S_k$:
because the trajectory visits each coordination environment $k$ in proportion to its probability $p_k$,
averaging along it weights each $S_k$ by $p_k$
without requiring $S_k$ to be computed for each environment separately.
Specifically,
the vibrational component may be evaluated from the phonon density of states
computed via the velocity autocorrelation function of the molecular dynamics trajectory,
and the electronic component from averaging the electronic entropy
using ab initio molecular dynamics along the same trajectory.
For single-component systems,
$k$ is specified by the number of nearest neighbors $N$ alone,
and the configurational entropy reduces to
\begin{equation}
    S_{\text{conf}} = -\sum_N p_N \ln p_N.
\end{equation}
For multicomponent systems, see Shang et al.\cite{Shang2025}

The separation of timescales is what makes this grouping useful:
vibrational sampling along a molecular dynamics trajectory
converges on accessible simulation timescales,
while the comparatively slow evolution of the coordination environment
contributes configurational entropy that is otherwise difficult to isolate.

\section{Conclusion}

We have presented a derivation of the partition function under coarse-graining
by maximizing entropy in its recursive form,
and argued that this form should be regarded as the primary expression for entropy.
We first derived the recursive property of entropy,
which separates entropy into configurational and intra-configurational contributions,
and showed that it reproduces the standard Gibbs entropy exactly
whenever each pure-state configuration belongs to exactly one coarse-grained configuration.
We then expressed the Helmholtz energy
in terms of intra-configurational Helmholtz energies $F_k$,
and maximizing entropy in recursive form gave the partition function $Z = \sum_k e^{-\beta F_k}$,
recovering Rushbrooke's result from the maximum-entropy principle.
Finally, we illustrated the usefulness of the framework
through two applications:
magnetic materials, where configurations are defined
by atomic spin arrangements,
and liquids, where configurations are defined
by nearest-neighbor environments.

Taken together, these results position zentropy as an exact framework
in which the scale of description is a free choice rather than an approximation.
Specifying a coarse-grained configuration constrains some degrees of freedom
and leaves the rest free.
The constrained degrees of freedom make approximations valid:
an approximation that would fail for the system as a whole
holds within a single configuration,
rendering otherwise intractable calculations tractable.
The free degrees of freedom carry the intra-configurational entropy $S_k$,
whose temperature dependence redistributes probability among configurations
and produces emergent behavior.

We anticipate that the present work
will help researchers understand what the zentropy approach does and why it works,
and will motivate members of the thermodynamics
and materials science communities
to utilize the recursive property and the zentropy approach as a framework
to render complex thermodynamic problems tractable,
particularly in density functional theory
and ab initio molecular dynamics,
but also wherever a system admits a natural hierarchical grouping.

\begin{acknowledgments}
This work is supported in part by the U.S. Department of Energy (DOE) under Grant No. DE-NE0009288, and in part by Penn State University through the 2025-2026 Opportunity Grant. This research was also partially supported by the Rising Researcher award ICDS\_RRYY\_027521 from Penn State’s ICDS (RRID:SCR\_025154).
\end{acknowledgments}

\section*{Data Availability}
No data were created or analyzed in this study.

\section*{CRediT author statement}

\noindent \textbf{Luke Allen Myers:} Conceptualization, Methodology, Writing - Original Draft, Writing - Review \& Editing, Visualization

\noindent \textbf{Nigel Lee En Hew:} Conceptualization, Writing - Review \& Editing, Supervision

\noindent \textbf{Shun-Li Shang:} Writing - Review \& Editing, Supervision, Funding acquisition

\noindent \textbf{Zi-Kui Liu:} Writing - Review \& Editing, Supervision, Funding acquisition

\bibliography{references}

@Article{Liu2019,
  author    = {Liu, Zi-Kui and Li, Bing and Lin, Henry},
  journal   = {J. Phase Equilib. Diff.},
  title     = {Multiscale Entropy and Its Implications to Critical Phenomena, Emergent Behaviors, and Information},
  year      = {2019},
  issn      = {1863-7345},
  month     = jun,
  number    = {4},
  pages     = {508--521},
  volume    = {40},
  doi       = {10.1007/s11669-019-00736-w},
  fjournal  = {Journal of Phase Equilibria and Diffusion},
  publisher = {Springer Science and Business Media LLC},
}

@Article{Wehrl1978,
  author    = {Wehrl, Alfred},
  journal   = {Rev. Mod. Phys.},
  title     = {General properties of entropy},
  year      = {1978},
  issn      = {0034-6861},
  month     = apr,
  number    = {2},
  pages     = {221--260},
  volume    = {50},
  doi       = {10.1103/revmodphys.50.221},
  fjournal  = {Reviews of Modern Physics},
  publisher = {American Physical Society (APS)},
}

@Book{Bengtsson2017,
  author    = {Bengtsson, Ingemar and Życzkowski, Karol},
  publisher = {Cambridge University Press},
  title     = {Geometry of Quantum States: An Introduction to Quantum Entanglement},
  year      = {2017},
  isbn      = {9781107656147},
  month     = aug,
  doi       = {10.1017/9781139207010},
}

@Article{Shannon1948,
  author    = {Shannon, C. E.},
  journal   = {Bell Syst. Tech. J.},
  title     = {A Mathematical Theory of Communication},
  year      = {1948},
  issn      = {0005-8580},
  month     = jul,
  number    = {3},
  pages     = {379--423},
  volume    = {27},
  doi       = {10.1002/j.1538-7305.1948.tb01338.x},
  fjournal  = {Bell System Technical Journal},
  publisher = {Institute of Electrical and Electronics Engineers (IEEE)},
}

@Book{Tolman1938,
  publisher = {Clarendon Press},
  address   = {Oxford},
  note      = {\url{https://archive.org/details/ThePrinciplesOfStatisticalMechanicsTolmanOxfordAtTheClarendonPress1938}},
  author = {Tolman, Richard C.},
  title  = {The Principles of Statistical Mechanics},
  year   = {1938},
  url    = {https://archive.org/details/ThePrinciplesOfStatisticalMechanicsTolmanOxfordAtTheClarendonPress1938},
}

@Book{Cover2005,
  author    = {Cover, Thomas M. and Thomas, Joy A.},
  publisher = {Wiley},
  title     = {Elements of Information Theory},
  year      = {2005},
  isbn      = {9780471748823},
  month     = apr,
  doi       = {10.1002/047174882x},
}

@Book{Nielsen2012,
  author    = {Nielsen, Michael A. and Chuang, Isaac L.},
  publisher = {Cambridge University Press},
  title     = {Quantum Computation and Quantum Information: 10th Anniversary Edition},
  year      = {2012},
  isbn      = {9780511976667},
  month     = jun,
  doi       = {10.1017/cbo9780511976667},
}

@Article{faddeev1956concept,
  author    = {Faddeev, Dmitrii Konstantinovich},
  journal   = {Uspekhi Mat. Nauk},
  title     = {On the concept of entropy of a finite probabilistic scheme},
  year      = {1956},
  number    = {1},
  pages     = {227--231},
  volume    = {11},
  fjournal  = {Uspekhi Matematicheskikh Nauk},
  publisher = {Russian Academy of Sciences, Steklov Mathematical Institute of Russian~…},
}

@Article{Aczel1974,
  author    = {Aczél, J. and Forte, B. and Ng, C. T.},
  journal   = {Adv. Appl. Probab.},
  title     = {Why the Shannon and Hartley entropies are ‘natural’},
  year      = {1974},
  issn      = {1475-6064},
  month     = mar,
  number    = {1},
  pages     = {131--146},
  volume    = {6},
  doi       = {10.2307/1426210},
  fjournal  = {Advances in Applied Probability},
  publisher = {Cambridge University Press (CUP)},
}

@Article{khinchin1953concept,
  author    = {Khinchin, Aleksandr Yakovlevich},
  journal   = {Uspekhi Mat. Nauk},
  title     = {The concept of entropy in the theory of probability},
  year      = {1953},
  number    = {3},
  pages     = {3--20},
  volume    = {8},
  fjournal  = {Uspekhi Matematicheskikh Nauk},
  publisher = {Russian Academy of Sciences, Steklov Mathematical Institute of Russian~…},
}

@Article{Jaynes1965,
  author    = {Jaynes, E. T.},
  journal   = {Am. J. Phys.},
  title     = {Gibbs vs Boltzmann Entropies},
  year      = {1965},
  issn      = {1943-2909},
  month     = may,
  number    = {5},
  pages     = {391--398},
  volume    = {33},
  doi       = {10.1119/1.1971557},
  fjournal  = {American Journal of Physics},
  publisher = {American Association of Physics Teachers (AAPT)},
}

@Article{Liu2024,
  author    = {Liu, Zi-Kui},
  journal   = {J. Phys. Condens. Matter},
  title     = {Quantitative predictive theories through integrating quantum, statistical, equilibrium, and nonequilibrium thermodynamics},
  year      = {2024},
  issn      = {1361-648X},
  month     = may,
  number    = {34},
  pages     = {343003},
  volume    = {36},
  doi       = {10.1088/1361-648x/ad4762},
  fjournal  = {Journal of Physics: Condensed Matter},
  publisher = {IOP Publishing},
}

@Article{Liu2024a,
  author    = {Liu, Zi-Kui and Hew, Nigel L. E. and Shang, Shun-Li},
  journal   = {Microstructures},
  title     = {Zentropy theory for accurate prediction of free energy, volume, and thermal expansion without fitting parameters},
  year      = {2024},
  issn      = {2770-2995},
  month     = jan,
  number    = {1},
  volume    = {4},
  doi       = {10.20517/microstructures.2023.56},
  publisher = {OAE Publishing Inc.},
}

@Article{Liu2022,
  author    = {Liu, Zi-Kui and Wang, Yi and Shang, Shun-Li},
  journal   = {J. Phase Equilib. Diff.},
  title     = {Zentropy Theory for Positive and Negative Thermal Expansion},
  year      = {2022},
  issn      = {1863-7345},
  month     = feb,
  number    = {6},
  pages     = {598--605},
  volume    = {43},
  doi       = {10.1007/s11669-022-00942-z},
  fjournal  = {Journal of Phase Equilibria and Diffusion},
  publisher = {Springer Science and Business Media LLC},
}

@Article{Bekenstein1975,
  author    = {Bekenstein, Jacob D.},
  journal   = {Phys. Rev. D},
  title     = {Statistical black-hole thermodynamics},
  year      = {1975},
  issn      = {0556-2821},
  month     = nov,
  number    = {10},
  pages     = {3077--3085},
  volume    = {12},
  doi       = {10.1103/physrevd.12.3077},
  fjournal  = {Physical Review D},
  publisher = {American Physical Society (APS)},
}

@Article{Rosas2024,
  journal   = {arXiv},
  pages     = {arXiv:2402.09090},
  author    = {Rosas, Fernando E. and Geiger, Bernhard C. and Luppi, Andrea I and Seth, Anil K. and Polani, Daniel and Gastpar, Michael and Mediano, Pedro A. M.},
  title     = {Software in the natural world: A computational approach to hierarchical emergence},
  year      = {2024},
  copyright = {arXiv.org perpetual, non-exclusive license},
  doi       = {10.48550/ARXIV.2402.09090},
  publisher = {arXiv},
}

@Article{Shang2023,
  author    = {Shang, Shun-Li and Wang, Yi and Liu, Zi-Kui},
  journal   = {Scripta Mater.},
  title     = {Quantifying the degree of disorder and associated phenomena in materials through zentropy: Illustrated with Invar Fe3Pt},
  year      = {2023},
  issn      = {1359-6462},
  month     = mar,
  pages     = {115164},
  volume    = {225},
  doi       = {10.1016/j.scriptamat.2022.115164},
  fjournal  = {Scripta Materialia},
  publisher = {Elsevier BV},
}

@Article{AdamGibbs1965,
  author   = {{Adam}, Gerold and {Gibbs}, Julian H.},
  journal  = {J. Chem. Phys.},
  title    = {On the Temperature Dependence of Cooperative Relaxation Properties in Glass-Forming Liquids},
  year     = {1965},
  number   = {1},
  pages    = {139--146},
  volume   = {43},
  adsurl   = {https://ui.adsabs.harvard.edu/abs/1965JChPh..43..139A},
  doi      = {10.1063/1.1696442},
  fjournal = {Journal of Chemical Physics},
}

@Article{MiracleSenkov2017,
  author   = {Miracle, Daniel B. and Senkov, Oleg N.},
  journal  = {Acta Mater.},
  title    = {A Critical Review of High Entropy Alloys and Related Concepts},
  year     = {2017},
  issn     = {1359-6454},
  pages    = {448--511},
  volume   = {122},
  doi      = {10.1016/j.actamat.2016.08.081},
  fjournal = {Acta Materialia},
  url      = {https://www.sciencedirect.com/science/article/pii/S1359645416306759},
}

@Article{Wang2025,
  journal   = {arXiv},
  pages     = {arXiv:2505.09851},
  author    = {Wang, Shun and Shang, Shun-Li and Liu, Zi-Kui and Hao, Wenrui},
  title     = {ZENN: A Thermodynamics-Inspired Computational Framework for Heterogeneous Data-Driven Modeling},
  year      = {2025},
  copyright = {Creative Commons Attribution 4.0 International},
  doi       = {10.48550/ARXIV.2505.09851},
  publisher = {arXiv},
}

@Article{Hew2024,
  author    = {Hew, Nigel Lee En and Shang, Shun-Li and Liu, Zi-Kui},
  journal   = {Phys. Rev. B},
  title     = {Predicting phase transitions in PbTiO3 using zentropy through quasiharmonic phonon calculations},
  year      = {2024},
  issn      = {2469-9969},
  month     = nov,
  number    = {18},
  pages     = {184103},
  volume    = {110},
  doi       = {10.1103/physrevb.110.184103},
  fjournal  = {Physical Review B},
  publisher = {American Physical Society (APS)},
}

@Article{Shang2010,
  author    = {Shang, Shun-Li and Saal, James E. and Mei, Zhi-Gang and Wang, Yi and Liu, Zi-Kui},
  journal   = {J. Appl. Phys.},
  title     = {Magnetic thermodynamics of fcc Ni from first-principles partition function approach},
  year      = {2010},
  issn      = {1089-7550},
  month     = dec,
  number    = {12},
  volume    = {108},
  doi       = {10.1063/1.3524480},
  fjournal  = {Journal of Applied Physics},
  publisher = {AIP Publishing},
}

@Article{Shang2010a,
  author    = {Shang, Shun-Li and Wang, Yi and Liu, Zi-Kui},
  journal   = {Phys. Rev. B},
  title     = {Thermodynamic fluctuations between magnetic states from first-principles phonon calculations: The case of bcc Fe},
  year      = {2010},
  issn      = {1550-235X},
  month     = jul,
  number    = {1},
  pages     = {014425},
  volume    = {82},
  doi       = {10.1103/physrevb.82.014425},
  fjournal  = {Physical Review B},
  publisher = {American Physical Society (APS)},
}

@Article{Shang2025,
  author    = {Shang, Shun-Li and Hew, Nigel L.E. and Gong, Rushi and Cockrell, Cillian and Bingham, Paul A. and Guo, Xiaofeng and Li, Jingjing and Hong, Qi-Jun and Liu, Zi-Kui},
  journal   = {J. Mol. Liq.},
  title     = {Achieving accurate entropy and melting point by ab initio molecular dynamics and zentropy theory: Application to fluoride and chloride molten salts},
  year      = {2025},
  issn      = {0167-7322},
  month     = nov,
  pages     = {128651},
  volume    = {438},
  doi       = {10.1016/j.molliq.2025.128651},
  fjournal  = {Journal of Molecular Liquids},
  publisher = {Elsevier BV},
}

@Article{Hong2025,
  author    = {Hong, Qi-Jun and Liu, Zi-Kui},
  journal   = {Phys. Rev. Research},
  title     = {Generalized approach for rapid entropy calculation of liquids and solids},
  year      = {2025},
  issn      = {2643-1564},
  month     = feb,
  number    = {1},
  pages     = {l012030},
  volume    = {7},
  doi       = {10.1103/physrevresearch.7.l012030},
  fjournal  = {Physical Review Research},
  publisher = {American Physical Society (APS)},
}

@Article{Wang2004,
  author    = {Wang, Y. and Liu, Z.-K. and Chen, L.-Q.},
  journal   = {Acta Mater.},
  title     = {Thermodynamic properties of Al, Ni, NiAl, and Ni3Al from first-principles calculations},
  year      = {2004},
  issn      = {1359-6454},
  month     = may,
  number    = {9},
  pages     = {2665--2671},
  volume    = {52},
  doi       = {10.1016/j.actamat.2004.02.014},
  fjournal  = {Acta Materialia},
  publisher = {Elsevier BV},
}

@Article{Csiszar2008,
  author    = {Csiszár, Imre},
  journal   = {Entropy-switz.},
  title     = {Axiomatic Characterizations of Information Measures},
  year      = {2008},
  issn      = {1099-4300},
  month     = sep,
  number    = {3},
  pages     = {261--273},
  volume    = {10},
  doi       = {10.3390/e10030261},
  fjournal  = {Entropy},
  publisher = {MDPI AG},
}

@Article{Rushbrooke1940,
  author    = {Rushbrooke, G. S.},
  journal   = {Trans. Faraday Soc.},
  title     = {On the statistical mechanics of assemblies whose energy-levels depend on the temperature},
  year      = {1940},
  issn      = {0014-7672},
  pages     = {1055},
  volume    = {36},
  doi       = {10.1039/tf9403601055},
  fjournal  = {Transactions of the Faraday Society},
  publisher = {Royal Society of Chemistry (RSC)},
}

@Article{Jaynes1957,
  author    = {Jaynes, E. T.},
  journal   = {Phys. Rev.},
  title     = {Information Theory and Statistical Mechanics},
  year      = {1957},
  issn      = {0031-899X},
  month     = may,
  number    = {4},
  pages     = {620--630},
  volume    = {106},
  doi       = {10.1103/physrev.106.620},
  fjournal  = {Physical Review},
  publisher = {American Physical Society (APS)},
}

@Article{Jaynes1957a,
  author    = {Jaynes, E. T.},
  journal   = {Phys. Rev.},
  title     = {Information Theory and Statistical Mechanics. II},
  year      = {1957},
  issn      = {0031-899X},
  month     = oct,
  number    = {2},
  pages     = {171--190},
  volume    = {108},
  doi       = {10.1103/physrev.108.171},
  fjournal  = {Physical Review},
  publisher = {American Physical Society (APS)},
}

@Article{Liu2015,
  author    = {Liu, Xuan L. and VanLeeuwen, Brian K. and Shang, Shun-Li and Du, Yong and Liu, Zi-Kui},
  journal   = {Nato. Sc. S. Ss. Iii. C. S.},
  title     = {On the scaling factor in Debye–Grüneisen model: A case study of the Mg–Zn binary system},
  year      = {2015},
  issn      = {0927-0256},
  month     = feb,
  pages     = {34--41},
  volume    = {98},
  doi       = {10.1016/j.commatsci.2014.10.056},
  fjournal  = {Computational Materials Science},
  publisher = {Elsevier BV},
}

@Article{Yuan2024a,
  author    = {Yuan, Kehua and Miao, Duoqian and Pedrycz, Witold and Ding, Weiping and Zhang, Hongyun},
  journal   = {IEEE Trans. Knowl. Data Eng.},
  title     = {Ze-HFS: Zentropy-Based Uncertainty Measure for Heterogeneous Feature Selection and Knowledge Discovery},
  year      = {2024},
  issn      = {2326-3865},
  month     = Nov,
  number    = {11},
  pages     = {7326--7339},
  volume    = {36},
  doi       = {10.1109/tkde.2024.3419215},
  fjournal  = {IEEE Transactions on Knowledge and Data Engineering},
  publisher = {Institute of Electrical and Electronics Engineers (IEEE)},
}

@Article{Hew2026,
  journal   = {arXiv},
  pages     = {arXiv:2604.17665},
  author    = {Hew, Nigel Lee En and Myers, Luke Allen and Shang, Shun-Li and Liu, Zi-Kui},
  title     = {pyzentropy: A Python package implementing recursive entropy for first-principles thermodynamics},
  year      = {2026},
  copyright = {Creative Commons Attribution 4.0 International},
  doi       = {10.48550/ARXIV.2604.17665},
  publisher = {arXiv},
}

@Article{Esposito2012,
  author    = {Esposito, Massimiliano},
  journal   = {Phys. Rev. E},
  title     = {Stochastic thermodynamics under coarse graining},
  year      = {2012},
  issn      = {1550-2376},
  month     = Apr,
  number    = {4},
  pages     = {041125},
  volume    = {85},
  doi       = {10.1103/physreve.85.041125},
  fjournal  = {Physical Review E},
  publisher = {American Physical Society (APS)},
}

@Article{Wang2008,
  author    = {Wang, Y. and Hector, L. G. and Zhang, H. and Shang, S. L. and Chen, L. Q. and Liu, Z. K.},
  journal   = {Phys. Rev. B},
  title     = {Thermodynamics of the {Ce} $\gamma$--$\alpha$ transition: Density-functional study},
  year      = {2008},
  issn      = {1550-235X},
  month     = sep,
  number    = {10},
  pages     = {104113},
  volume    = {78},
  doi       = {10.1103/physrevb.78.104113},
  fjournal  = {Physical Review B},
  publisher = {American Physical Society (APS)},
}

@Article{Shore1980,
  author    = {Shore, J. and Johnson, R.},
  journal   = {IEEE Trans. Inf. Theory},
  title     = {Axiomatic derivation of the principle of maximum entropy and the principle of minimum cross-entropy},
  year      = {1980},
  issn      = {1557-9654},
  month     = Jan,
  number    = {1},
  pages     = {26--37},
  volume    = {26},
  doi       = {10.1109/tit.1980.1056144},
  fjournal  = {IEEE Transactions on Information Theory},
  publisher = {Institute of Electrical and Electronics Engineers (IEEE)},
}

@Article{Noid2008,
  author    = {Noid, W. G. and Chu, Jhih-Wei and Ayton, Gary S. and Krishna, Vinod and Izvekov, Sergei and Voth, Gregory A. and Das, Avisek and Andersen, Hans C.},
  journal   = {J. Chem. Phys.},
  title     = {The multiscale coarse-graining method. I. A rigorous bridge between atomistic and coarse-grained models},
  year      = {2008},
  issn      = {1089-7690},
  month     = jun,
  number    = {24},
  volume    = {128},
  doi       = {10.1063/1.2938860},
  fjournal  = {The Journal of Chemical Physics},
  publisher = {AIP Publishing},
}

@Article{Stillinger1982,
  author    = {Stillinger, Frank H. and Weber, Thomas A.},
  journal   = {Phys. Rev. A},
  title     = {Hidden structure in liquids},
  year      = {1982},
  issn      = {0556-2791},
  month     = Feb,
  number    = {2},
  pages     = {978--989},
  volume    = {25},
  doi       = {10.1103/physreva.25.978},
  fjournal  = {Physical Review A},
  publisher = {American Physical Society (APS)},
}

@Article{Ceder1993,
  author    = {Ceder, G.},
  journal   = {Nato. Sc. S. Ss. Iii. C. S.},
  title     = {A derivation of the Ising model for the computation of phase diagrams},
  year      = {1993},
  issn      = {0927-0256},
  month     = Mar,
  number    = {2},
  pages     = {144--150},
  volume    = {1},
  doi       = {10.1016/0927-0256(93)90005-8},
  fjournal  = {Computational Materials Science},
  publisher = {Elsevier BV},
}

@Article{Lee2012,
  author    = {Lee, Julian and Pressé, Steve},
  journal   = {Phys. Rev. E},
  title     = {Microcanonical origin of the maximum entropy principle for open systems},
  year      = {2012},
  issn      = {1550-2376},
  month     = Oct,
  number    = {4},
  pages     = {041126},
  volume    = {86},
  doi       = {10.1103/physreve.86.041126},
  fjournal  = {Physical Review E},
  publisher = {American Physical Society (APS)},
}

@Book{Eddington1928,
  author    = {Eddington, Arthur Stanley},
  publisher = {Cambridge University Press},
  title     = {The Nature of the Physical World},
  year      = {1928},
  address   = {Cambridge},
}

% \appendix
% \renewcommand\thefigure{\thesection.\arabic{figure}}
% \setcounter{figure}{0}
% \renewcommand{\thetable}{A\arabic{table}}
% \setcounter{table}{0}
% \input{appendix}

\end{document}